# Multi-material topology optimization of adhesive backing layers via J-integral and strain energy minimizations

**Zhiyuan Tong**

Department of Mechanical Engineering, University of British Columbia, Vancouver, BC V6T 1Z4, Canada

**Farid H. Benvidi**

Department of Mechanical Engineering, University of British Columbia, Vancouver, BC V6T 1Z4, Canada

**Mattia Bacca**[1]

Department of Mechanical Engineering, University of British Columbia, Vancouver, BC V6T 1Z4, Canada e-mail: mbacca@mech.ubc.ca

[1] Corresponding author

Strong adhesives often rely on reduced stress concentrations obtained via specific functional grading of material properties. This can be seen in many examples in nature and engineering. Basic design principles have been formulated based on parametric optimization, but a general design tool is still missing. We propose here the use of topology optimization to achieve optimal stiffness distribution in a multi-material adhesive backing layer, reducing stress concentration at selected (crack tip) locations. The method involves the minimization of a linear combination of (i) the J-integral around the crack tip and (ii) the strain energy of the structure. This combination is due to the compromise between numerical stability and accuracy of the method, where (i) alone is numerically unstable and (ii) alone cannot eliminate the crack-tip stress singularity. We analyze three cases in plane strain conditions, namely (1) double-edged crack and (2) center crack, in tension, and (3) edge crack under shear. Each case evidences a different optimal topology with (1-2) providing similar results. The optimal topology allocates stiffness in regions that are far away from the crack tip, and the allocation of softer materials over stiffer ones produces a sophisticated structural hierarchy. To test our solutions, we plot the contact stress distribution across the interface. In all observed cases, we eliminate the stress singularity at the crack tip, albeit generating (mild) stress concentrations in other locations. The optimal topologies are tested to be independent of crack size. Our method ultimately provides the robust design of flaw tolerant adhesives where the crack location is known.

## 1. Introduction

Adhesion is a complex chemo-mechanical problem, and is ubiquitous in engineering and nature. Strong adhesives commonly rely on a multi-material composition, whereby the inhomogeneous spatial distribution of properties, *i.e.* stiffness, can enhance adhesive strength [1-5]. This is due to the contact stress distribution, where stress concentrations are responsible for the initiation and propagation of interfacial flaws. The interplay between geometry and mechanics in this case suggests the need for constructing powerful design tools.

A notable example of geometry-driven adhesion enhancement is in the development of bioinspired fibrillar adhesion [6-7], where the contact is split into multiple units (contact splitting) to reduce stress concentrations at the interface, and thus develop robust adhesion. Moreover, to increase adhesion, the geometry of the single unit has shown to play an important role [8-9], where mushroom-shaped fibrils have proved to be far superior then pillar-like fibrils [9-12], and a quest for the optimal shape has begun [13-18]. In the same context, composite fibrils having a soft tip have shown stronger adhesion compared to single-material ones [3-4,19], and researchers have attempted again to achieve optimal design by defining the optimal shape of the bi-material interface [20-22]. This analysis, however, consider only two materials and are mainly based on the optimization of specific parameter values, where certain features of the optimal geometry are known at priory. To define the optimal design of an adhesive, however, one would require a method involving a generic number of materials, and where the geometry of the adhesive is completely unknown.

In this paper, we propose the use of topology optimization to explore the best distribution of material rigidity (stiffness) in the backing layer of the adhesive to enhance adhesive strength. The simplest and most traditional method is the Solid Isotropic Material with Penalization (SIMP) [23-24]. It has been initially developed for structural optimization to design lightweight structure by locating material where mostly needed and leave structural gaps where the presence of material is less relevant for the overall load bearing. The resulting optimal structure depends on the geometry of the design domain and the boundary conditions (loading and constraints). Structural optimization has also been used in the context of fracture [25-29] to minimize the driving force for crack propagation. In this case, unlike adhesion, the path of crack propagation can be unpredictable, thus [30,28-29] proposed the adoption of the phase field method to construct a reliable prediction. In the context of adhesion, [31] utilized similar methods to design the optimal interfacial topology to resist delamination of mono-dimensional and bi-dimensional structures. In the context of optimal design of adhesives, [32] proposed a non-parametric (topology based) model for the design of lap joints based on strain energy minimization. They consider the possibility of a multi-material optimization and finally obtain a reduction in stress concentration. However, the latter is a side product of strain energy minimization, and not a direct minimization of stress concentrations. An alternative approach is the direct minimization of stress concentrations with methods such as the stress-constrained optimization [28-29,33-35]. However, one has to identify the stress component to be minimized. Similarly, [36] proposed a method that combines strain energy minimization with the minimization of the nonuniformity of contact pressure, in frictionless conditions. Adhesion problems, however, involve the concentration of both contact pressure and shear stress, *i.e.*, mixed-mode fracture. Thus, the hypothesis of frictionless conditions is not applicable in adhesion, particularly in the case of lap shears. This gives merit to methods based on the minimization of (scalar) energetic quantities such as the energy release rate, *i.e.*, the J-integral [37]. Moreover, [36] does not consider multi-material optimization. Only by adopting a multi-material method one can map the hierarchy of needs in terms of material rigidity within the structure.

Our model system consists in a design domain composed of multiple materials. Here, we minimize a linear combination of the strain energy of the adhesive backing, and the J-integral computed around the tip of a pre-defined crack. The specification of the crack location is necessary to utilize J-integral minimization, which ensures minimal strain energy release rate. Hence, the proposed model is best suited for problems where detachment starts at known locations. An example is the case where debonding initiates where the adhesive is exposed to external agents, *i.e.* stress corrosion cracking of adhesives [38]. As will be explained in the next sections, strain energy minimization alone can result in slightly reducing stress concentrations, and thus produce a slight increase in adhesion, but cannot completely eliminate the stress singularity at the tip of the crack. J-integral minimization alone, on the other hand, produces a computational error, which will be explained in the *Discussion and Conclusions* section. The combined minimization of both the above quantities ultimately produces the best results. We utilize a blending parameter $\alpha$, where $\alpha = 0$ gives strain energy minimization only, and $\alpha = 1$ only gives J-integral minimization. As $\alpha$ increases, the optimal adhesive reduces the J-integral, and the adhesive becomes stronger. Above an intermediate value of $\alpha$, the topology becomes stationary with $\alpha$ and the J-integral reduces only slightly, suggesting an optimal tradeoff. Also, an increase in $\alpha$ can aggravate stress concentrations in other locations than the crack tip. The use of multi-material optimization proves to be valuable in defining the hierarchy of needs for structural rigidity in the topology of the adhesive. I.e. as shown in the results section, using, e.g. three materials with different moduli, the optimal topology will allocate the stiffest material where rigidity is most valuable and the softest material where rigidity is suboptimal. Finally, we explore the effect of Poisson's ratio and the number of materials in the final topology and contact stress distribution.

## 2. Topology Optimization Method

The design domain is composed of $m$ materials distributed within the domain via the functions $\rho_i(x)$ describing the nominal density of material $i$, with $i = 1, ..., m$, and $x = (x, y)$ the position within the domain $V$. The functions $\rho_i(x)$ define then the 'topology' of the structure, and are subject to the conditions

$$\rho_i(x) \in [0,1], \forall i = 1, ..., m \tag{1a}$$

$$\sum_{i=1}^{m} \rho_i(x) \leq 1, \forall x \in V \tag{1b}$$

$$f_i = f_{0i}, \forall i = 1, ..., m \tag{1c}$$

with

$$f_i = \frac{1}{V}\int_V \rho_i(x) dV \tag{1d}$$

where $f_i$ and $f_{0i}$ are the volume fraction of material $i$ in the whole domain, and its prescribed value, respectively. By summing Eq. (1c) over $i$, from (1b), we can obtain $\sum_{i=1}^{m} f_i \leq 1$. For the case in which $\sum_{i=1}^{m} f_i < 1$, a soft matrix occupies the portion of domain that is unoccupied by the $m$ materials. *Material-blending* giving $\rho_i > 0$ or $\rho_i < 1$ provides unrealistic topology, and is thus penalized in our method as will be explained later.

The displacement field $u(x)$ generates the strain tensor $\varepsilon(x) = (\nabla u + \nabla u^T)/2$, and the elastic stress tensor $\sigma(x)$, given by Hooke's law as

$$\sigma = \frac{\nu E}{(1+\nu)(1-2\nu)}\varepsilon_v I + \frac{E}{(1+\nu)}\varepsilon \tag{2}$$

Here, $\varepsilon_v$ is the volumetric strain, i.e. the trace of $\varepsilon$, $I$ is the identity matrix, and $E$ and $\nu$ are the Young modulus and the Poisson's ratio of the material. The elastic constants $E$ and $\nu$ depend on the material composition, hence on the constants of the $m$ materials and their density. The stresses $\sigma$ must be equilibrated with respect to external forces, while the displacement field $u$ must be compatible with the prescribed constraints.

The topology of the structure is considered optimal if it minimizes the objective function $\phi$. Our method considers a combination of the J-integral, $J$, and the strain energy of the structure, $C$, also called 'compliance', giving

$$\phi = \alpha \frac{J}{J_0} + (1-\alpha)\frac{C}{C_0} \tag{3a}$$

where $\alpha$ is a dimensionless parameter defining the combination, and $J_0$ and $C_0$ are the nominal J-integral and compliance, calculated for the homogeneous, unoptimized, structure. Here, the J-intgreal is [39]

$$J = -\int_\Omega \left(w \frac{\partial s}{\partial x} - \frac{\partial u^T}{\partial x} \sigma \nabla s\right) d\Omega \tag{3b}$$

with $\Omega$ the integration domain indicated in Figure 1 touching the crack faces,

$$w = \frac{1}{2}\sigma \cdot \varepsilon \tag{3c}$$

the strain energy density of the material, · the scalar product, and $\nabla s$ the gradient of the (smooth) test function $s$, satisfying the condition of $s = 1$ on the inner boundary of $\Omega$ and $s = 0$ on the outer boundary of $\Omega$. In Figure 1, we report the contour plot of the function $s$, and the spider mesh with quarter point elements used around the crack tip. These elements are specifically designed to capture the stress singularity around the crack tip.

The compliance is

$$C = \int_V w dV \tag{3d}$$

and it constitutes the traditional objective function for structural optimization. While minimization of $J$ would ensure the reduction of stress concentrations at the crack tip (giving stronger adhesion) as elaborated in the Discussion and Conclusions section, J-integral alone cannot be used as objective function due to the non-monotonic relation between $J$ and the material density $\rho_i$. Moreover, in some cases the simple use of $C$ minimization ensures the reduction of stress concentration (see the Results section).

The minimization of $\phi$, in Eq. (3), under the conditions at Eq. (1) and (2) involves the stationarity of the Lagrange function

$$\mathcal{L} = \phi + \lambda_i(f_i - f_{0i}) \tag{4}$$

Due to the imposed conditions at Eq. (1c), we have $\mathcal{L} = \phi$. Hence minimization of $\mathcal{L}$ provides the minimization of $\phi$.

The proposed model utilizes the Solid Isotropic Material with Penalty (SIMP) method, based on finite element analysis (FEA). This method is described in the following sections.

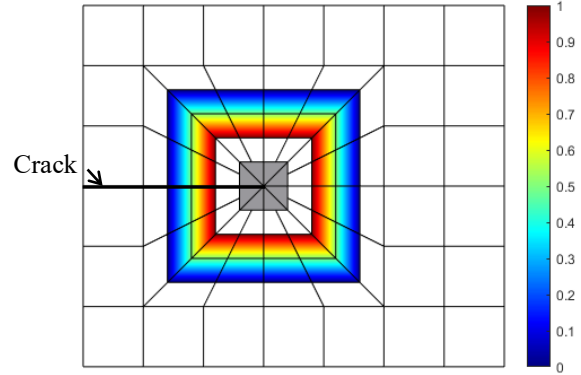

*Figure 1*: Contour plot of the test function $s$ within the domain $\Omega$, and spider mesh with quarter point elements around the crack tip.

### 2.1. Density update

Via FEA discretization, the topology of the structure is now described by $\rho_{ie}$, the density of material $i$ in element $e$, with $e = 1, ..., n$ and $n$ the number of elements. Minimization of $\mathcal{L}$ is obtained by imposing its stationarity with respect to $\rho_{ie}$ and $\lambda_i$. The latter simply enforces Eq. (1c) while the former gives that

$$\frac{\partial \mathcal{L}}{\partial \rho_{ie}} = \frac{\partial \phi}{\partial \rho_{ie}} + \lambda_i A_e \tag{5}$$

must be zero, with $A_e$ the area of element $e$. The density update follows then with

$$\rho_{ie,new} = \rho_{ie} r_{ie}^d \tag{6a}$$

where

$$r_{ie} = -\frac{1}{\lambda_i A_e} \frac{\partial \phi}{\partial \rho_{ie}} \tag{6b}$$

is the update ratio and $d < 1$ is a damping coefficient to avoid excessive jumps in the value of $\rho_{ie}$ that would produce numerical instabilities. Here we adopt $d = 0.3$, as it has proven to be an effective value in providing stable results with a relatively low number of iterations. From Eq. (6) it can be deduced that, when $r_{ie} > 1$, $\partial \mathcal{L}/\partial \rho_{ie} < 0$ so that an increment in $\rho_{ie}$ produces a reduction of $\mathcal{L}$, thus favoring its minimization. The sensitivity of the objective function, $\partial \phi/\partial \rho_{ie}$, is calculated in the next section.

## 2.2. Sensitivity analysis

FEA involves mapping of $\boldsymbol{u}$ as $\boldsymbol{u}(\boldsymbol{x}) = \boldsymbol{N}(\boldsymbol{x})\boldsymbol{U}$, where $\boldsymbol{N}(\boldsymbol{x})$ is the shape-function tensor, and the vector $\boldsymbol{U}$ contains the nodal values of $\boldsymbol{u}$. For planar problems, e.g. plane strain or plane stress, the strain vector $\hat{\boldsymbol{\varepsilon}} = (\varepsilon_x, \varepsilon_y, 2\varepsilon_{xy})$ is

$$\hat{\boldsymbol{\varepsilon}} = \boldsymbol{B}\,\boldsymbol{U} \tag{7a}$$

and the displacement gradient, transposed, is

$$\frac{\partial \boldsymbol{u}^T}{\partial x} = \breve{\boldsymbol{B}}\,\boldsymbol{U} \tag{7b}$$

with the tensors $\boldsymbol{B}$ and $\breve{\boldsymbol{B}}$ containing the components of the gradient of $\boldsymbol{N}$. By rewriting Eq. (3) for our palanar problem, the compliance becomes

$$C = \tfrac{1}{2} \boldsymbol{U}^T \boldsymbol{K} \boldsymbol{U} \tag{8a}$$

with

$$\boldsymbol{K} = \int_A \boldsymbol{B}^T \boldsymbol{E} \boldsymbol{B} dA \tag{8b}$$

the stiffness matrix of the structure, and $\boldsymbol{E}$ the elastic tensor giving $\hat{\boldsymbol{\sigma}} = \boldsymbol{E}\,\hat{\boldsymbol{\varepsilon}}$, where $\hat{\boldsymbol{\sigma}} = (\sigma_x, \sigma_y, \sigma_{xy})$. In the same way, the J-integral becomes

$$J = \tfrac{1}{2} \boldsymbol{U}^T (\boldsymbol{M}_1 - \boldsymbol{M}_2) \boldsymbol{U} \tag{9a}$$

with

$$\boldsymbol{M}_1 = 2 \int_\Omega \breve{\boldsymbol{B}}^T \begin{pmatrix} \frac{\partial s}{\partial x} & 0 & \frac{\partial s}{\partial y} \\ 0 & \frac{\partial s}{\partial y} & \frac{\partial s}{\partial x} \end{pmatrix} \boldsymbol{E} \boldsymbol{B} d\Omega \tag{9b}$$

and

$$\boldsymbol{M}_2 = \int_\Omega \boldsymbol{B}^T \frac{\partial s}{\partial x} \boldsymbol{E} \boldsymbol{B} d\Omega \tag{9c}$$

From Eq. (3), (8) and (9) we can then write

$$\phi = \tfrac{1}{2} \boldsymbol{U}^T \boldsymbol{M} \boldsymbol{U} \tag{10a}$$

with

$$\boldsymbol{M} = \frac{\alpha}{J_0}(\boldsymbol{M}_1 - \boldsymbol{M}_2) + \frac{1-\alpha}{C_0} \boldsymbol{K} \tag{10b}$$

The sensitivity becomes then

$$\frac{\partial \phi}{\partial \rho_{ie}} = \tfrac{1}{2} \boldsymbol{U}^T (\boldsymbol{M} + \boldsymbol{M}^T) \frac{\partial \boldsymbol{U}}{\partial \rho_{ie}} + \tfrac{1}{2} \boldsymbol{U}^T \frac{\partial \boldsymbol{M}}{\partial \rho_{ie}} \boldsymbol{U} \tag{11}$$

Here, given that equilibrium imposes stationarity of $C$, via

$$\boldsymbol{U} = \boldsymbol{K}^{-1} \boldsymbol{F} \tag{12}$$

with $\boldsymbol{F}$ the (constant) vector of nodal forces, and $d\boldsymbol{K}^{-1} = -\boldsymbol{K}^{-1} d\boldsymbol{K} \boldsymbol{K}^{-1}$, we have that

$$\frac{\partial \boldsymbol{U}}{\partial \rho_{ie}} = -\boldsymbol{K}^{-1} \frac{\partial \boldsymbol{K}}{\partial \rho_{ie}} \boldsymbol{U} \tag{13}$$

By substituting Eq. (13) into (11) we finally have

$$\frac{\partial \phi}{\partial \rho_{ie}} = \tfrac{1}{2} \boldsymbol{U}^T \left[ -(\boldsymbol{M} + \boldsymbol{M}^T) \boldsymbol{K}^{-1} \frac{\partial \boldsymbol{K}}{\partial \rho_{ie}} + \frac{\partial \boldsymbol{M}}{\partial \rho_{ie}} \right] \boldsymbol{U} \tag{14}$$

The density $\rho_{ie}$ only affects the rigidity of the material in element $e$ by incrementing the amount of material $i$ in it. We assume that each material has the same $\nu$, while only $E$ varies within the domain. Considering also the linear correlation between the matrices $\boldsymbol{M}$ and $\boldsymbol{K}$ with $E$, we can rewrite the derivatives in Eq. (14) as

$$\frac{\partial \boldsymbol{K}}{\partial \rho_{ie}} = \frac{\partial \boldsymbol{K}}{\partial E} \frac{\partial E}{\partial \rho_{ie}} \tag{15a}$$

$$\frac{\partial \boldsymbol{M}}{\partial \rho_{ie}} = \frac{\partial \boldsymbol{M}}{\partial E} \frac{\partial E}{\partial \rho_{ie}} \tag{15b}$$

where the term $\partial E/\partial \rho_{ie}$ is calculated from the relation $E = E(\rho_{ie})$ given in the next section.

## 2.3. Elastic constants and material-blending penalty

The elastic constant of element $e$ is given by

$$E = c_e E_e \tag{16a}$$

where

$$E_e = E_0 + \sum_{i=1}^{m}(E_i - E_0) \rho_{ie}^p \tag{16b}$$

with $E_0$ the modulus of the soft matrix, and $p > 1$ a penalty coefficient. We adopt $p = 3$. In Eq. (16a), we also have

$$c_e = 1 + c_0 - c_1 \left(\sum_{i=1}^{m} \rho_{ie}\right)^q \tag{16c}$$

with $c_0$ a small but non-zero coefficient (we take $c_0 = 10^{-5}$),

$$c_1 = 1 - \frac{1}{2(m-1)} \sum_{i=1}^{m} \sum_{j=1}^{m} \left[ \left(\rho_{ie} - \rho_{je}\right)^2 \right]^r \tag{16d}$$

and $q > 1$ and $r \geq 1$ penalty coefficients. Here we adopt $q = 5$ and $r = 1$.

The above penalty method is constructed to provide the highest penalty for the case of material blending, i.e. when element $e$ is composed by more than just one material or the soft matrix. This can be deduced by calculating $\partial E/\partial \rho_{ie}$ from Eq. (16).

In Eq. (15) we have then

$$\frac{1}{E}\frac{\partial E}{\partial \rho_{ie}} = \frac{1}{c_e}\frac{\partial c_e}{\partial \rho_{ie}} + \frac{1}{E_e}\frac{\partial E_e}{\partial \rho_{ie}} \tag{17a}$$

with

$$\frac{\partial c_e}{\partial \rho_{ie}} = -\frac{\partial c_1}{\partial \rho_{ie}}\left(\sum_{j=1}^{m} \rho_{je}\right)^q - c_1 q \rho_{ie}^{q-1} \tag{17b}$$

$$\frac{\partial c_1}{\partial \rho_{ie}} = -\frac{2r}{m-1} \sum_{j=1}^{m} \left(\left(\rho_{ie} - \rho_{je}\right)^2\right)^{r-1} \left(\rho_{ie} - \rho_{je}\right) \tag{17c}$$

and

$$\frac{\partial E_e}{\partial \rho_{ie}} = p(E_i - E_0)\rho_{ie}^{p-1} \qquad (17d)$$

*2.4. Iterations and convergence*

By substituting Eq. (17) into (15), the result into (14), then into (6), we can update the variable $\rho_{ie}$ at each step, but that requires the calculation of the Lagrange multiplier $\lambda_i$ satisfying Eq. (1c). This is performed via bisection method, starting from a tentative value for $\lambda_i$.

The stability of the method and the convergence of $\rho_{ie}$ toward the optimum distribution is ensured if $r_{ie} > 0$ and $r_{ie}$ converges to unity. To ensure $r_{ie} > 0$, we need $\partial \phi / \partial \rho_{ie} < 0$, from Eq. (6b).

From Eq. (3a), the sensitivity rewrites as

$$\frac{\partial \phi}{\partial \rho_{ie}} = \frac{\alpha}{J_0} \frac{\partial J}{\partial \rho_{ie}} + \frac{1-\alpha}{C_0} \frac{\partial C}{\partial \rho_{ie}} \qquad (18)$$

Here $\partial C / \partial \rho_{ie}$ is always negative, and this can be seen by rearranging Eq. (8a), from equilibrium at Eq. (12), to

$$C = \frac{1}{2} \mathbf{F}^T \mathbf{K}^{-1} \mathbf{F} \qquad (19)$$

where an increment of $\rho_{ie}$ raises the rigidity of the material and hence reduces the compliance $C$ (assuming the force vector $\mathbf{F}$ constant). However, $\partial J / \partial \rho_{ie}$ can be positive since an increase in the rigidity of the material might increase the stress concentration around the crack tip. For this reason, when $\alpha = 1$, convergence is not ensured and our method can generate unreliable results. In the proposed method, we adopt the constraint $\partial \phi / \partial \rho_{ie} \leq 0$, so that positive values of sensitivity are taken as zero to ensure stability. This has shown to ensure the convergence of the method for $\alpha < 1$.

## 3. Results

In this section, we report the main results obtained with the proposed method. All our results consider plane strain hypothesis, since this is the most representative for most adhesives. A simple extension of the model can provide results for axisymmetric adhesives, and is lest for future exploration.

In Figure 2, we show the legend of the notations used in this section (left) and the schematics of the problem setting (right). Here, the design domain constitutes the adhesive backing, sandwiched between the bottom (adhesive) layer and the top layer, with respective thicknesses $h_b$ and $h_t$. The adhesive layer is in adhesive contact with the target surface lying below it (not shown in the figure). The adhesive is debonded from the target surface at one if its edges, by a length $l_{crack}$, which we treat as a crack (contact defect) propagating in mixed-mode. Adopting the size of the adhesive backing as unity, we have $W = H = 1$, $l_{crack} = h_b = 0.025$, and $h_t = 0.25$. From this, we have that $l_{crack}/W \ll 1$, i.e. the crack has infinitesimal size compared to the adhesive backing. We also have that $h_b = l_{crack}$, where the thickness of the adhesive layer is comparable with the crack length, so that the stresses transmitted in the design domain preserve the concentration near the crack tip.

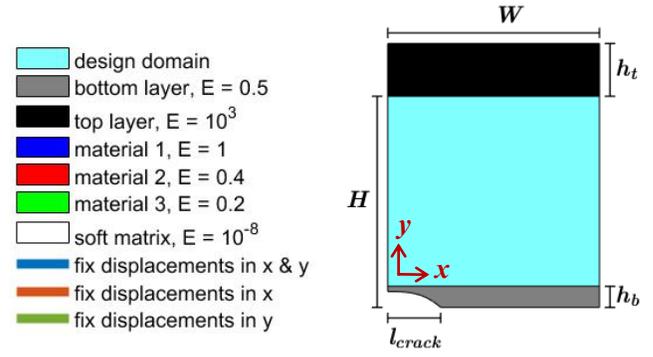

*Figure 2*: Legend of the notations used in this section (left), and schematics of the problem setting (right), where $W = H = 1$, $h_b = l_{crack} = 0.025$, and $h_t = 0.25$. The volume fractions of materials are $f_1 = 0.2$, and $f_2 = f_3 = 0.1$.

The backing of the adhesive is made of three materials, i.e. $m = 3$. We order them from the stiffest to the softest, with material 1 (the stiffest) having unit modulus. As shown in Figure 2 (left), we have $E_1 = 1$, then $E_2 = 0.4$ $E_3 = 0.2$, and the soft matrix has modulus $E_0 = 10^{-8}$. The volume fractions of materials are $f_1 = 0.2$, and $f_2 = f_3 = 0.1$. By virtue of linear elasticity, the stresses calculated in the analysis should be considered as normalized by $E_1$. The modulus of the adhesive layer is $E_b = 0.5$, hence intermediate to $E_1$ and $E_0$. This arbitrary choice has shown not to affect the results qualitatively, as $E_b$ simply modulates the stress concentration transmitted to the backing. We consider the top layer to be rigid, adopting $E_t = 10^3$. The top layer is commonly stiffer than the adhesive (e.g. structural materials have moduli ~ GPa, while polymer-based adhesive have modulus ~ MPa).

The load applied is a uniform normal stress for Case 1 and 2 and uniform shear stress for Case 3. In all cases the stress value is $P = 0.005$. This value corresponds to 0.5% the nominal modulus $E_1$, and is chosen to be small for consistency with the hypothesis of linear elasticity

Figure 3 reports the three cases analyzed. From left to right: case 1, with normal loading $P$ and double-edged crack; case 2, with normal loading and center crack; case 3, with shear loading. In the latter case, we position the crack so that the loading produces tension at the crack tip. In cases 1 and 2, the right edge is a plane of symmetry. For all three cases, we assume plane strain deformation.

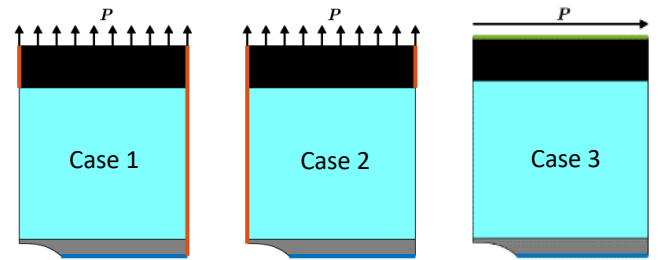

*Figure 3*: We analyze three cases: case 1 (left) with normal load and double-edge crack; case 2 (center) with normal load and center crack; case 3 with shear load. The notation is reported in Figure 2.

For all cases, we will analyze the evolution of the optimized structure for various values of $\alpha$, investigate the distribution of contact stresses, and analyze the influence of the Poisson's ratio $\nu$ and number of materials $m$. In all cases, $l_{crack}$ has shown not to affect the results so long that $l_{crack} \ll H, W$.

## 3.1. Case 1

Figure 4 reports the computed optimal topology for Case 1 (normal loading and double-edged crack, as in Figure 3 left), for $\nu = 0.45$ and $\alpha = 0$, 0.2 and 0.7. When $\alpha = 0$, the method only reduces compliance, thereby only providing load transmission from top to bottom with minimum strain energy. This involves allocation of material near the crack tip, hence promoting stress concentration. For larger $\alpha$, the material is allocated at the center of the design domain, away from the crack tip. The color map indicates where the stiffer material (blue), the intermediate material (red) and the softer material (green) are needed.

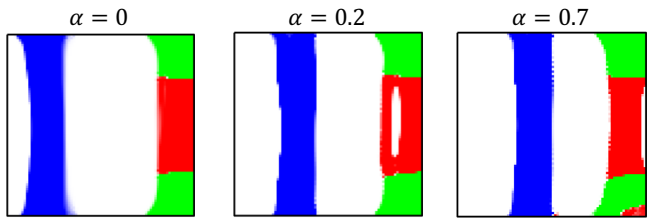

*Figure 4*: Computed optimal topology for Case 1 (Figure 4 left) with $\nu = 0.45$, and $\alpha = 0$, 0.2 and 0.7.

Figure 5 shows the optimal values for $C/C_0$ and $J/J_0$ at different $\alpha$. Here one can deduce that $\alpha > 0$ reduces the J-integral up to $\alpha = 0.7$. For $\alpha > 0.7$ we do not observe a significant improvement, hence we omit these results. In fact, from Figure 4 we observe a modest structural evolution from $\alpha = 0.2$ to $\alpha = 0.7$, and in Figure 5 we see that the reduction in $J$ becomes less significant around $\alpha = 0.7$.

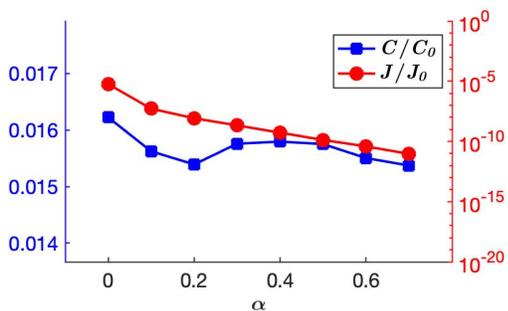

*Figure 5*: Optimal dimensionless compliance $C/C_0$ and J-integral $J/J_0$ versus $\alpha$ for the cases analyzed in Figure 4.

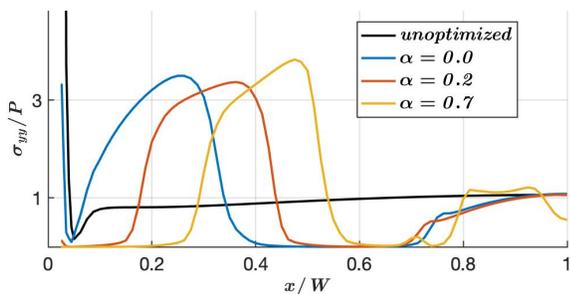

*Figure 6*: Distribution of contact stresses $\sigma_{yy}$ versus $x$ for the cases analyzed in Figure 4.

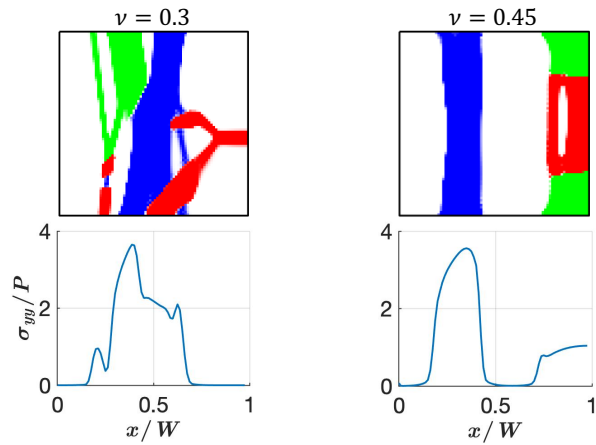

*Figure 7*: Computed optimal topology (top) and contact stress (bottom) for Case 1 (Figure 3) with $\nu = 0.3$ and 0.45, for $\alpha = 0.3$

Figure 6 shows the distribution of dimensionless contact stress $\sigma_{yy}/P$ for the unoptimized (homogeneous) structure, as well as for the optimal structure with $\alpha = 0$, $\alpha = 0.2$, and $\alpha = 0.7$. The homogeneous structure presents a stress singularity at the crack tip, as commonly predicted from linear elastic fracture mechanics. For $\alpha = 0$, the singularity is still present but significantly reduced. As the value of $\alpha$ increases to higher values, the singularity reduces until it completely vanishes for the case of $\alpha = 0.7$. For all the three cases analyzed here, we observe stress concentrations appearing at the right (internal) edge and at intermediate regions within the interface. For highest $\alpha$, the right edge concentration reduces while the intermediate concentration increases. This was also observed in composite adhesives having a rigid stalk and a thin soft tip [20]. Our method proves then to be valid in increasing adhesive strength against edge detachment. This is the case for adhesives exposed to aggressive environment where the edges in contact with external agents are more likely to initiate a crack [38]. However, in adhesives where detachment can occur at any location in the contact, our method provides useful results for $\alpha = 0$, but should be extended to account for a random location of the crack.

In Figure 7 we analyze the influence of the Poisson's ratio $\nu$, showing the optimized structure (top) and the contact stress distribution (bottom), for $\alpha = 0.3$. The structure produced by $\nu = 0.3$ is more complex than that produced by $\nu = 0.45$. The intermediate stress concentrations are comparable among the two examples, with $\nu = 0.45$ producing a more staggered structure.

In Figure 8 we analyze the influence of the number of materials, showing the optimized structure (top) and the distribution of contact stresses (bottom), for $\alpha = 0.3$. We preserve the volume fraction of the soft matrix in all three analyses. Thus, for $m = 1$, we have $f_1 = 0.4$; for $m = 2$, we have $f_1 = f_2 = 0.2$; and for $m = 3$, we have $f_1 = 0.2$ and $f_2 = f_3 = 0.1$. As can be observed from Figure 8, the structure becomes more complex and slenderer (staggered) as $m$ (number of materials) increases from 1 to 3 while keeping the overall structural profile for $m = 1,2$ and departing slightly for $m = 3$. This provides an overview of the hierarchy of needs for rigidity within the structure. With an increasing number of materials, the stress singularity reduces until it disappears for three materials. On the other hand, the stress concentration in other locations increase with the number of materials.

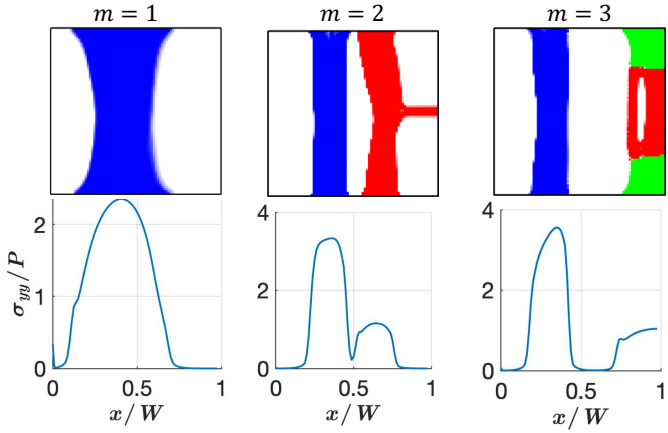

*Figure 8*: Computed optimal topology (top) and contact stress (bottom) for Case 1 (Figure 3) with $\nu = 0.45$ and for $\alpha = 0.3$, for $m = 1,2,3$. The material volume fractions are $f_1 = 0.4$ for $m = 1$, $f_1 = f_2 = 0.2$ for $m = 2$, and $f_1 = 0.2$ and $f_2 = f_3 = 0.1$ for $m = 3$

### 3.2. Case 2

Figure 9 reports the computed optimal topology for Case 2 (normal loading and center crack), for $\nu = 0.45$ and various $\alpha$. Like in the previous case, higher $\alpha$ relocates the material far from the crack tip to reduce its stress singularity.

Figure 10 shows the optimal values for $C/C_0$ and $J/J_0$ at different $\alpha$. Here we can see that higher $\alpha$ reduces again the J-integral as in Case 1.

Figure 11 shows the distribution of dimensionless contact normal stresses $\sigma_{yy}/P$ for the homogeneous structure (unoptimized) as well as for cases analyzed above. For $\alpha = 0$ we can see a reduced stress singularity at the crack tip on the left, while for $\alpha > 0$ this disappears. At the (right) edge of the adhesive, we have an edge stress singularity (defect free) for the homogeneous structure, and this reduces significantly for all the analyzed optimal topologies. This concentration, however, becomes more severe for increasing $\alpha$ due to the relocation of material from left to right. Because the stress singularity at the center (left edge of the reported half-specimen) is eliminated with $\alpha$ as low as 0.2, higher $\alpha$ provide limited benefits and instead produces higher concentration at the (right) edge.

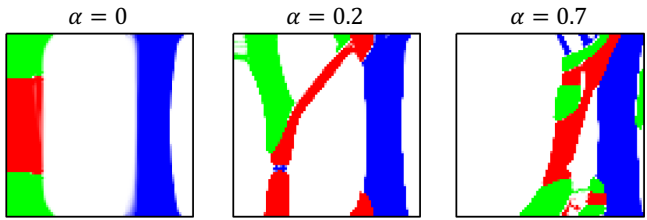

*Figure 9*: Computed optimal topology for case 2 (Figure 4 center) with $\nu = 0.45$, and $\alpha = 0, 0.2$ and $0.7$.

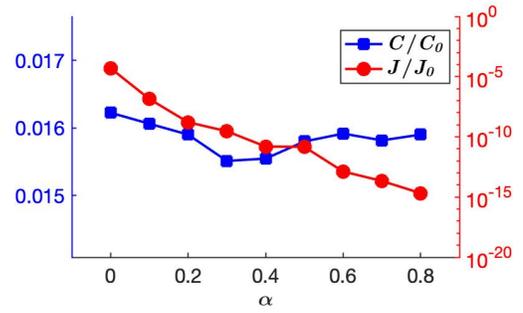

*Figure 10*: Optimal dimensionless compliance $C/C_0$ and J-integral $J/J_0$ versus $\alpha$ for the cases analyzed in Figure 7.

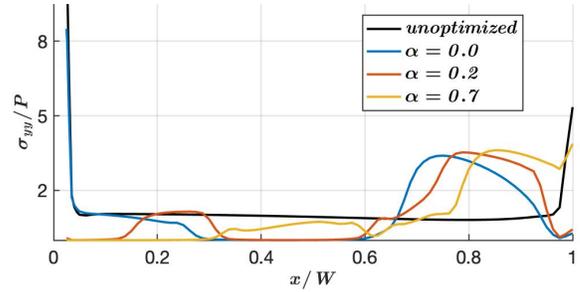

*Figure 11*: Distribution of contact stresses $\sigma_{yy}$ versus $x$ for the cases analyzed in Figure 7.

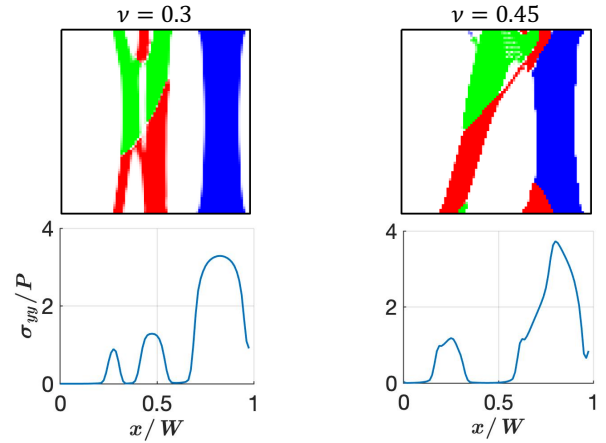

*Figure 12*: Computed optimal topology (top) and contact stress (bottom) for Case 2 (Figure 3) with $\nu = 0.3$ and $0.45$, for $\alpha = 0.3$

In Figure 12 we analyze the influence of the Poisson's ratio $\nu$, showing the optimized structure (top) and the contact stress distribution (bottom), for $\alpha = 0.3$. The structure produced by $\nu = 0.3$ is again more complex than that produced by $\nu = 0.45$, with the intermediate stress concentrations being comparable among the two examples. In this case, $\nu = 0.3$ produces a more staggered structure than $\nu = 0.45$.

In Figure 13 we analyze the influence of the number of materials, showing the optimized structure (top) and the distribution of contact stresses (bottom), for $\alpha = 0.3$. The volume fractions of materials are the same as in Figure 8. The structure becomes more complex and slenderer as $m$ (number of materials) increases from 1 to 3, while preserving the overall profile. As a consequence of the increased complexity, the stress concentration at the right edge, as well as intermediate stress concentrations, become more severe with more materials.

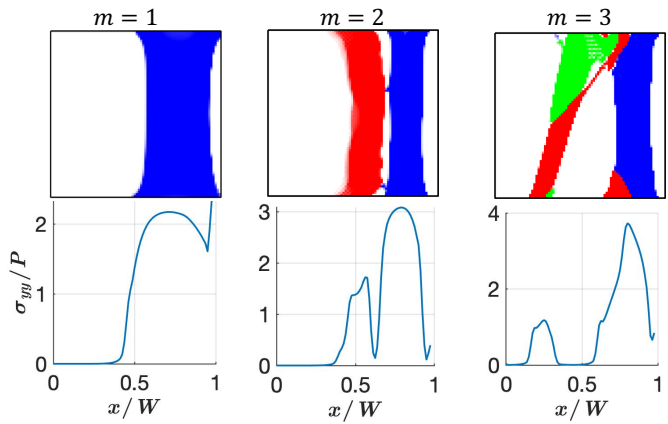

*Figure 13*: Computed optimal topology (top) and contact stress (bottom) for Case 2 (Figure 3) with $v = 0.45$, for one, two and three materials, *i.e.* for $m = 1,2,3$. The volume fractions of materials are the same as in Figure 8.

### 3.3. Case 3

Figure 14 reports the computed optimal topology for case 3 (shear loading, as in Figure 3 right), for $v = 0.45$ and various $\alpha$. Here, the value of $\alpha$ appears to have a more reduced impact on the results compared with the previous cases. This is likely due to the reduced stress concentration at the crack tip due to a higher mode mixing. Also, for $\alpha = 0$ the structure is symmetric while for higher $\alpha$ the location of the crack influences the optimal topology, thereby disrupting the symmetry.

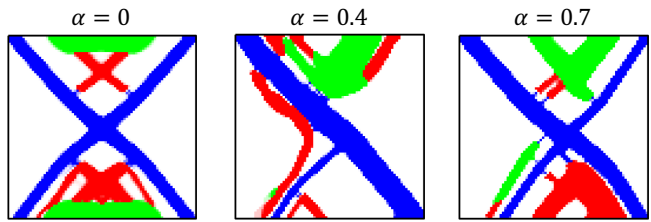

*Figure 14*: Computed optimal topology for case 3 (Figure 4 right) with $v = 0.45$, and $\alpha = 0$, 0.2 and 0.4.

Figure 15 shows the optimal values for $C/C_0$ and $J/J_0$ at different $\alpha$. Here we can see that for $0.1 < \alpha < 0.8$ $J/J_0$ is nearly constant, with a slight further reduction for $\alpha = 0.9$.

Figure 16 shows the distribution of dimensionless contact normal stresses $\sigma_{yy}/P$ for the unoptimized structure and for case analyzed above. For $\alpha = 0$ we can see the stress concentration at the crack tip on the left edge, while for $\alpha > 0$ this disappears. For all values of $\alpha$, the right edge shows a compressive stress concentration (negative $\sigma_{yy}$), which increases with $\alpha$. Compressive stress does not prompt detachment, but instead help promoting better interfacial conformation to the surface roughness of the adhered material, and hence strengthen adhesion on the right edge. For simplicity, we ignore compressive stress concentrations. It should be noted that the optimal structure observed for $\alpha > 0$ is asymmetric. Thus, this optimality is applicable only for the case in which the direction of the shear load and the location of the edge crack is defined according to the chosen configuration.

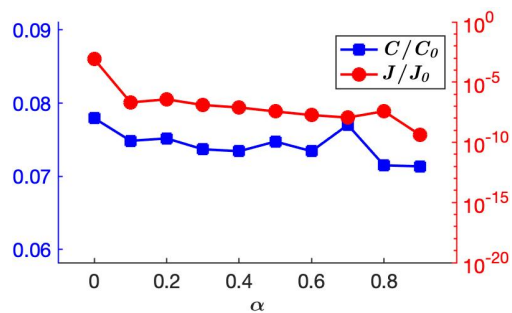

*Figure 15*: Optimal dimensionless compliance $C/C_0$ and J-integral $J/J_0$ versus $\alpha$ for the cases analyzed in Figure 10.

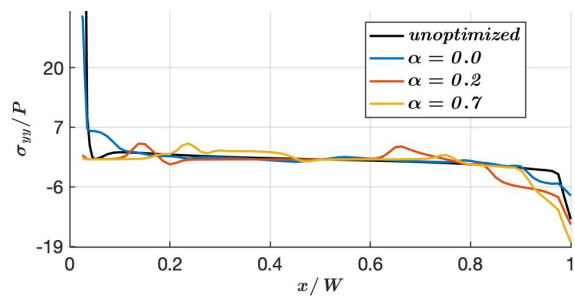

*Figure 16*: Distribution of contact stresses $\sigma_{yy}$ versus $x$ for the cases analyzed in Figure 10.

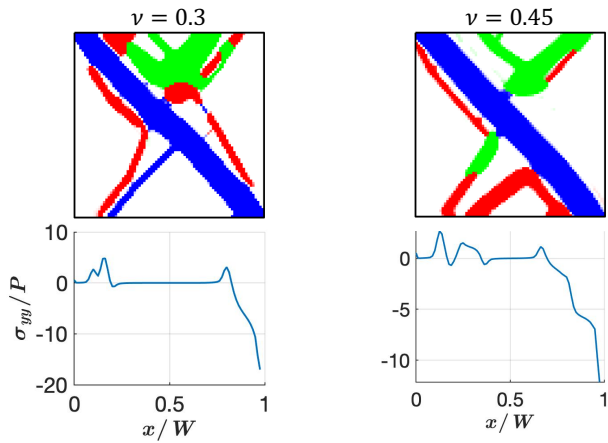

*Figure 17*: Computed optimal topology (top) and contact stress (bottom) for Case 2 (Figure 3) with $v = 0.3$ and 0.45, for $\alpha = 0.3$

In Figure 17 we analyze the influence of the Poisson's ratio $v$, showing the optimized structure (top) and the contact stress distribution (bottom), for $\alpha = 0.3$. The structural complexity in this case is comparable, and so is the contact stress distribution.

In Figure 18 we analyze the influence of the number of materials, showing the optimized structure (top) and the distribution of contact stresses (bottom), for $\alpha = 0.3$. The volume fractions of materials are the same as in Figure 8. The structure becomes slightly more complex and slenderer as $m$ (number of materials) increases while keeping the overall profile. Both structural features and contact stress distribution appear to change only slightly with $m$, and we can again observe the hierarchy of needs in structural rigidity. In all cases we can observe an asymmetric topology, suggesting again the specificity in the load direction and crack location.

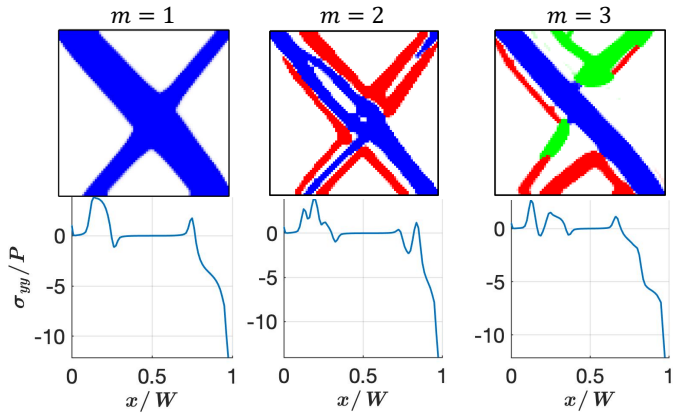

*Figure 18*: Computed optimal topology (top) and contact stress (bottom) for Case 3 (Figure 3) with $v = 0.45$, for one, two and three materials, *i.e.* for $m = 1,2,3$. The volume fractions of materials is the same as in Figure 8.

**Discussion and Conclusions**

The results obtained with $\alpha = 0$, *i.e.* by compliance minimization only, evidence a reduced stress singularity at the crack tip, compared with the homogeneous adhesive (unoptimized). This proves that strain energy minimization alone can promote adhesive strength. This is simply due to the fact that the stress singularity around the crack tip also creates a strain energy accumulation in that region. Apparently, a reduction of the global strain energy produces a reduction in stress concentration at the crack tip. As we observed, however, strain energy minimization alone cannot remove the stress singularity, hence one has to also minimize the J-integral around the crack tip. A SIMP-based topology optimization algorithm like the one proposed does not ensure stability and convergence with J-integral minimization alone. This is due to the fact that the J-integral does not correlate monotonically with the modulus of the material, unlike compliance (and strain energy), while our method requires that the sensitivity ($\partial \phi / \partial \rho_{ie}$) remains negative. To overcome this limitation, we truncate to zero the positive sensitivity. This has proven to promote convergence for the proposed combined minimization of compliance and J-integral, for $\alpha < 1$. For small $\alpha$, as $\alpha$ increases, the J-integral decreases by several orders of magnitude, while for higher $\alpha$ the results show more modest improvement. This is also because, at higher $\alpha$, this variable produces small changes in the topology of the adhesive. In conclusion, the free parameter $\alpha$ produces some uncertainty, but above a sufficiently large $\alpha$, the results do not change qualitatively. The threshold value in $\alpha$ depends on the specific case of study, where for Case 3, we have $\alpha_{th} \sim 0.1$, while Cases 1-2 show $\alpha_{th} \sim 0.5$.

To generate physically realistic results, our method penalizes material mixing so that the final solution will provide a specific material and not a mixing of multiple materials in each point of the domain of the structure. This, however, can produce unrealistic checkerboard topology in subdomains of the structure. To prevent this numerical artifact from affecting our results, our code includes a checkerboard filter based on the homogenization of the local sensitivity.

J-integral minimization is effective when detachment propagates from known locations such as the edges of the adhesive (Case 1,3) and the center (Case 2,3). Edge crack propagation is mostly observed when debonding is promoted by environmental attack, due to the exposure of the edge to chemical contaminants. Center crack propagation is observed as an alternative mechanism in homogeneous adhesives, due to interfacial stress concentrations [20]. However, as observed in Figures 6-8, 11-13, and 16-18, the optimized structure in our study shows interfacial stress concentrations away from edge and center. This limits the applicability of our method to the case of detachment by propagation statistical interfacial defects, and one should extend our model to consider the simultaneous presence of multiple cracks randomly located across the interface. Linear elastic fracture mechanics provides the J-integral estimation $J = \Theta l_{crack} w$, with $\Theta$ a geometric coefficient that accounts for the crack location and size ($\Theta \approx 2\pi$ for infinitesimal edge cracks), $l_{crack}$ the crack length, and $w = P^2/2E'$ an effective strain energy density, where $P$ is the remotely applied stress, and $E' = E/(1 - v^2)$ the effective plane strain modulus of the material releasing strain energy upon crack propagation. In the observed optimized structures, $J$ reduces by several orders of magnitude, compared to the case of homogeneous backing layer. This improvement scales with the contrast in modulus $E'$ between the soft matrix and the stiff inclusions. Crack propagation stability occurs under the condition $\partial J/\partial l_{crack} < 0$, associated with a reduction of $\Theta$ and/or an increment in $E'$ as the crack propagates. Functional grading of material modulus, like in the explored case, is expected to exhibit this phenomenon and provide a crack trapping mechanism, as observed in the case of fibrillar adhesives [40-41]. However, this is left for future development.

The Poisson's ratio has significant influence on the optimal topology. Typical adhesives are made of rubbery polymers, hence incompressible. We adopted $v = 0.45$, as the case of near-incompressibility, since higher $v$ produced numerical instabilities.

We also analyzed the case of different crack length and observed negligible difference in the results. This is because the optimal topology creates a load transmission path that is directed away from the crack tip to reduce stress concentration. As a result, the optimal adhesive is flaw insensitive, rendering it suitable for practical applications, where the crack length is unknown at priori.

Finally, the use of multiple materials allows us to identify the spatial resolution of rigidity requirements to reach the optimal structure. For Case 1 and 2, where the external loads are a uniform normal stress at the top of the adhesive, the optimal topology requires more rigidity near the external edge, while softer materials are commonly located near the center of the adhesive.

**Acknowledgments**

This work is financially supported by the Natural Sciences and Engineering Research Council of Canada (NSERC) (RGPIN-2017), and by the Human Frontiers in Science Program (RGY0073/2020).

**References**

[1] H. Peisker, J. Michels, S.N. Gorb (2013), "Evidence for a material gradient in the adhesive tarsal setae of the ladybird beetle Coccinella septempunctata," *Nature Communication* **4**:1661

[2] H.K. Minsky, K.T. Turner (2015), "Achieving enhanced and tunable adhesion via composite posts," *Applied Physics Letters* **106**(20):201604.

[3] S.C.L. Fischer, E. Arzt, R. Hensel (2017), "Composite Pillars With a Tunable Interface for Adhesion to Rough Substrates," *ACS Applied Materials and Interfaces*, **9**(1):1036-1044

[4] L. Heepe, S. Höft, J. Michels, S.N. Gorb (2018), "Material Gradients in Fibrillar Insect Attachment Systems: The Role of Joint-Like Elements," *Soft Matter*, **14**(34):7026-7033

[5] S. Kumar, B.L. Wardle, M.F. Arif, J. Ubaid (2018), "Stress Reduction of 3D Printed Compliance-Tailored Multilayers," *Advanced Engineering Materials* **20**:1700883

[6] N.J. Glassmaker, A. Jagota, C.Y. Hui, J. Kim (2004), "Design of biomimetic fibrillar interfaces: 1. Making contact," *Journal of the Royal Society Interface*, **123**(33): 1700883

[7] H. Gao, X. Wang, H. Yao, S. Gorb, E. Arzt (2005), "Mechanics of hierarchical adhesion structures of geckos," *Mechanics of Materials*, **37**(2-3):275-285

[8] A. Del Campo, C. Greiner, E. Arzt (2007), "Contact shape controls adhesion of bioinspired fibrillar surfaces," *Langmuir*, **23**(20):10235-10243.

[9] A.V. Spuskanyuk, R.M. McMeeking, V.S. Deshpande, E. Arzt (2008), "The effect of shape on the adhesion of fibrillar surfaces," *Acta biomaterialia*, **4**(6):1669-1676.

[10] G. Carbone, E. Pierro, S.N. Gorb (2011), "Origin of the superior adhesive performance of mushroom-shaped microstructured surfaces," *Soft Matter*, **7**(12):5545-5552

[11] C.K. Hossfeld, A.S. Schneider, E. Arzt, C.P. Frick (2013),17 "Detachment behavior of mushroom-shaped fibrillar adhesive surfaces in peel testing," *Langmuir*, **29**(49):15394-15404

[12] H. Marvi, S. Song, M. Sitti (2015), "Experimental investigation of optimal adhesion of mushroomlike elastomer microfibrillar adhesives," *Langmuir*, **31**(37):10119-10124

[13] G. Carbone, E. Pierro (2012), "Sticky Bio-inspired Micropillars: Finding the Best Shape," *Small*, **8**(9):1449-1454

[14] B. Aksak, K. Sahin, M. Sitti (2014), "The optimal shape of elastomer mushroom-like fibers for high and robust adhesion," *Beilstein journal of nanotechnology* **5**(1):630-638

[15] M. Micciché, E. Arzt, E. Kroner (2014), "Single macroscopic pillars as model system for bioinspired adhesives: influence of tip dimension, aspect ratio, and tilt angle," *ACS applied materials & interfaces*, **6**(10):7076-7083

[16] Y. Kim, C. Yang, Y. Kim, G.X. Gu, S. Ryu (2020), "Designing an adhesive pillar shape with deep learning-based optimization," *ACS applied materials & interfaces*, **12**(21):24458-24465

[17] D. Son, V. Liimatainen, M. Sitti (2021), "Machine Learning‐Based and Experimentally Validated Optimal Adhesive Fibril Designs," *Small*, **17**(39):2102867

[18] A. Kong, M. Bacca (2022), "A self-adhesion criterion for slanted micropillars," *Extreme Mechanics Letters*, **52**:101663

[19] R.G. Balijepalli, S.C. Fischer, R. Hensel, R.M. McMeeking, E. Arzt (2017), "Numerical study of adhesion enhancement by composite fibrils with soft tip layers," *Journal of the Mechanics and Physics of Solids*, **99**:357-378

[20] F.H. Benvidi, M. Bacca (2021), "Theoretical Limits in Detachment Strength for Axisymmetric Bi-Material Adhesives," *Journal of Applied Mechanics*, **88**(12): 121007

[21] A. Luo, K.T. Turner (2021), "Achieving enhanced adhesion through optimal stress distributions," *Journal of the Mechanics and Physics of Solids*, **156**:104610

[22] A. Luo, H. Zhang, K.T. Turner (2022), "Machine learning-based optimization of the design of composite pillars for dry adhesives," *Extreme Mechanics Letters*, **54**:101695

[23] M.P. Bendsøe (1989), "Optimal shape design as a material distribution problem," *Structural optimization*, **1**(4):193-202

[24] M.P. Bendsøe, O. Sigmund (1999), "Material interpolation schemes in topology optimization," *Archive of applied mechanics*, **69**(9):635-654

[25] P. Chaperon, R. Jones, M. Heller, S. Pitt, F. Rose (2000), "A methodology for structural optimisation with damage tolerance constraints," *Engineering Failure Analysis*, **7**(4):281-300

[26] V.J. Challis, A.P. Roberts, A.H. Wilkins (2008), "Fracture resistance via topology optimization," *Structural & Multidisciplinary Optimization*, **36**:263–271

[27] T.T. Banh, N.G. Luu, D. Lee (2021), "A non-homogeneous multi-material topology optimization approach for functionally graded structures with cracks," *Composite Structures*, **273**:114230

[28] J.B. Russ, H. Waisman (2019). Topology optimization for brittle fracture resistance. *Computer Methods in Applied Mechanics and Engineering*, **347**:238-263

[29] J.B. Russ, H. Waisman (2020). A novel topology optimization formulation for enhancing fracture resistance with a single quasi-brittle material. *International Journal for Numerical Methods in Engineering*, **121**:2827-2856

[30] L. Xia, D. Da, J. Yvonnet (2018). Topology optimization for maximizing the fracture resistance of quasi-brittle composites. *Computer Methods in Applied Mechanics and Engineering*, **332**:234-254

[31] K. Sylves, K. Maute, M.L. Dunn (2009), "Adhesive surface design using topology optimization," *Structural and Multidisciplinary Optimization*, **38**(5):455-468

[32] H. Ejaz, A. Mubashar, I.A. Ashcroft, E. Uddin, M. Khan (2018), "Topology optimisation of adhesive joints using non-parametric methods," *International Journal of Adhesion and Adhesives*, **81**:1-10

[33] P. Duysinx, M.P. Bendsøe (1998). Topology optimization of continuum structures with local stress constraints. International Journal of Numerical Methods for Engineering, **43**:1453-1478

[34] G. Cheng, Z. Jiang (1992). Study on Topology Optimization with Stress Constraints. *Engineering Optimization*, **20**(2):129-148

[35] C. Le, J. Norato, T. Bruns, C. Ha, D. Tortorelli (2010). Stress-based topology optimization for continua. *Structural and Multidisciplinary Optimization*, **41**:605-620

[36] C. Niu, W. Zhang, T. Gao (2019). Topology optimization of continuum structures for the uniformity of contact pressures. *Structural and Multidisciplinary Optimization*, **60**:185–210

[37] Z. Kang, P. Liu, M. Li (2017). Topology optimization considering fracture mechanics behaviors at specified locations. *Structural and Multidisciplinary Optimization*, **55**:1847-1864

[38] A. Leronni, N.A. Fleck (2022), "Delamination of a sandwich layer by diffusion of a corrosive species: Initiation of growth," *Journal of the Mechanics and Physics of Solids*, **160**:104775

[39] J.R. Rice (1968). A path independent integral and the approximate analysis of strain concentration by notches and cracks.

[40] M. Bacca, J.A. Booth, K. Turner, R.M. McMeeking (2016). Load sharing in bioinspired fibrillar adhesives with backing layer interactions and interfacial misalignment. *Journal of the Mechanics and Physics of Solids*, **96**: 428–444

[41] H. Khungura, M. Bacca (2020). Optimal load sharing in bioinspired fibrillar adhesives: Asymptotic solution. *Journal of Applied Mechanics*, **88**(3):031004